\documentclass[aps,prd,preprintnumbers,nofootinbib,preprint,showkeys]{revtex4}
\usepackage{bm}
\usepackage{latexsym}
\usepackage{dcolumn}
\usepackage{amsmath,amsfonts,amssymb}
\usepackage{graphicx}
\usepackage{color}
\usepackage{amsthm}
\usepackage{epstopdf}
\usepackage{hyperref}

\begin{document}
\title{Black Hole with Quantum Potential}

\author{Ahmed Farag Ali$^{1}$}\email[]{ahmed.ali@fsc.bu.edu.eg}
\author{Mohammed M. Khalil$^{2}$} \email[]{moh.m.khalil@gmail.com}
\affiliation{$^1$Department of Physics, Faculty of Science, Benha University, Benha 13518, Egypt}
\affiliation{$^2$Department of Electrical Engineering,\\ Alexandria University,  Alexandria 12544, Egypt}

\begin{abstract}
In this work, we investigate black hole (BH) physics in the context of quantum corrections. These quantum corrections were introduced recently by replacing classical geodesics with quantal (Bohmian) trajectories and hence form a quantum Raychaudhuri equation (QRE). From the QRE, we derive a modified Schwarzschild metric, and use that metric to investigate BH singularity and thermodynamics. We find that these quantum corrections change the picture of Hawking radiation greatly when the size of BH approaches the Planck scale. They prevent the BH from total evaporation, predicting the existence of a quantum BH remnant, which may introduce a possible resolution for the catastrophic behavior of Hawking radiation as the BH mass approaches zero. Those corrections also turn the spacelike singularity of the black hole to be timelike, and hence this may ameliorate the information loss problem. 
\end{abstract}

\keywords{quantum aspects of black holes; black hole singularity; Raychaudhuri equation; modified theories of gravity}

\maketitle

\section{Introduction}

Recently, a new semi-classical approach for quantum gravity has been suggested in \cite{Das:2013oda}, in which it was shown that replacing classical trajectories, or geodesics, by  quantal (Bohmian) trajectories leads to corrections to the Raychaudhuri equation.
Hence, these new quantum corrections will affect all reasonable spacetimes which  are incomplete or singular in a certain sense depending on the validity of the classical Raychaudhuri equation according to Hawking-Penrose theorem \cite{Hawking:1969sw}. 
The quantum Raychaudhuri equation (QRE) has been found to prevent focusing of geodesics, and hence prevents the formation of singularities \cite{Das:2013oda}. This has been investigated in cosmology with Friedmann-Robertson-Walker (FRW) Universe and it was found that the big bang singularity may be resolved using quantal geodesics \cite{Ali:2014qla}. It was also found that the Friedmann equation receives a quantum correction term that could be interpreted as a cosmological constant that gives a correct estimate of its observed value \cite{Ali:2014qla, Das:2014agf}.

The QRE has been derived in \cite{Das:2013oda} by considering a quantum mechanical description of a fluid or condensate. This condensate is described by a wavefunction $\psi={\cal R}e^{iS}$, which is assumed to be normalizable and single valued,
${\cal R} (x^a)$ and $S (x^a)$ are real functions associated with the four velocity field $u_a = (\hbar/m) \partial_a S$. The expansion scalar is given by $\theta=Tr(u_{a;b}) = h^{ab} u_{a;b}$, where the transverse metric $h_{ab}=g_{ab}+u_au_b$.
The quantum Raychaudhuri equation for timelike geodesics, with vanishing shear and twist for simplicity, takes the form after some derivations \cite{Das:2013oda, Ali:2014qla}
\begin{equation}
\frac{d\theta}{d\lambda} =- \frac{1}{3}~\theta^2
- R_{ab} u^a u^b  +  \frac{\hbar^2}{m^2} h^{ab} \left( \frac{\Box {\cal R}}{\cal R} \right)_{;a;b}
+ \frac{\epsilon_1 \hbar^2}{m^2} h^{ab} R_{;a;b} ~.
\label{QREtime} 
\end{equation}
$R_{a,b}$ and $R$ are the Ricci tensor and Ricci scalar respectively. The constant $\epsilon_1=1/6$ for conformally invariant scalar fluid, but left arbitrary here.

Since the black hole is an ideal laboratory to investigate quantum gravity, in this paper, we derive the quantum Raychaudhuri equation for null geodesics, and use it to derive a modified Schwarzschild metric. We then derive the quantum corrected thermodynamics of the black hole. We also 
investigate the impact of quantum corrections on the physical singularity of the black hole.

\section{Quantum Raychaudhuri Equation for Null Geodesics}

In the case of null geodesics, the classical Raychaudhuri equation with vanishing shear and twist takes the form \cite[P. 50]{poisson2004}
\begin{equation}
	\frac{d\theta}{d\lambda}=-\frac{1}{2}\theta^2-R_{ab}k^ak^b,
\end{equation}
where $k^a$ is a null tangent vector field and the expansion parameter $\theta$ is given by
\begin{equation}
	\theta=k^a_{~;a}.
\end{equation}
The transverse metric $h_{ab}$ is given by \cite[P.46]{poisson2004}
\begin{equation}
\label{auxmetric}
h_{ab} = g_{ab}+ k_a N_b + N_a k_b,
\end{equation}
where $N_a$ is an auxiliary null  vector field such that $k^aN_a=-1$ and $N^aN_a=0$. To derive the quantum Raychaudhuri equation for null geodesics, we start with a Klein-Gordon-type equation with $m=0$ 
\begin{equation}
\label{KGeq}
\left( \Box  -\epsilon_1 R -\epsilon_2 \frac{i}{2} f_{cd}\sigma^{cd} \right)\Phi =0,
\end{equation}
where $R$ is the curvature scalar, 
and $\epsilon_1 = 1/6$ for conformally invariant scalar field. The 4-momentum and ``coordinate velocity'' are defined as \cite{Sasaki,horton,durr}
\begin{align}
	& p_a =\hbar\partial_a S, \\
	& \vec v = \frac{d\vec x}{dt} = - c^2 \frac{\vec\nabla S}{\partial^0 S},
\end{align}
where we used the fact that the null vector $k_a$ is defined up to a constant \cite[P. 86]{hawking1973}, and since $k_a$ is any null tangent vector, it is appropriate that one chooses this to be the momentum vector $p_a=\hbar k_a$. Substituting the wave function $\psi(\vec x,t)={\cal R} e^{iS}$ in Eq.\eqref{KGeq} yields the two equations
\begin{align}
&\partial^a \left( {\cal R}^2 \partial_a S \right) = \frac{\epsilon_2}{2}~f_{cd} \sigma^{cd} {\cal R}^2,  \label{kgcont} \\
& p^2 = \hbar^2 \epsilon_1 R -  \hbar^2\frac{\Box {\cal R}}{\cal R}.
\label{kgeom}
\end{align}
where Eq.\eqref{kgeom} yields
the modified geodesic equation with the relativistic quantum potential term $V_Q=\hbar^2 \frac{\Box {\cal R}}{\cal R}$
\begin{equation}
p^b_{~;a} p^a =  \hbar^\epsilon_1 R^{;b} -
\hbar^2\left( \frac{\Box {\cal R}}{\cal R} \right)^{;b}~.
\end{equation}
Then the quantum corrected Raychaudhuri equation for null geodesics takes the form
\begin{equation}
\frac{d\theta}{d\lambda} =  - \frac{1}{2}~\theta^2	- R_{ab} k^a k^b   + \epsilon_1 \hbar^2 h^{ab} R_{;a;b} + \hbar^2 h^{ab} \left( \frac{\Box {\cal R}}{\cal R} \right)_{;a;b}~.
\label{qre2}
\end{equation}
This equation is useful for studying black holes, since the event horizon of the black hole is a null surface generated by null geodesics \cite{poisson2004}.

\section{Quantum-Modified Schwarzschild Metric}
\label{sec:metric}

To derive the Schwarzschild metric, we start from the general metric for a static spherically symmetric spacetime
\begin{equation}
ds^2=-\alpha(r)dt^2+\beta(r)dr^2+r^2 d\Omega,
\label{metric}
\end{equation}
where $\alpha(r)$ and $\beta(r)$ are functions of $r$ alone and go to one at infinity. Since the metric is independent of $t$, then there exists a Killing vector $K^a=(1,0,0,0)$. Hence, along a geodesic with affine parameter $\lambda$ we have
\begin{equation}
K_a\frac{dx^a}{d\lambda}=\text{constant}.
\end{equation}
For the metric \eqref{metric} this leads to 
\begin{equation}
\label{atlambda}
\alpha \frac{dt}{d\lambda}=\text{constant}.
\end{equation}
On a radial outgoing null geodesic $u=t-\int\beta(r)dr$ is constant, and
\begin{equation}
\label{ka}
k_a=-\partial_a u
\end{equation}
is a null tangent vector field with $+r$ as an affine parameter \cite[P. 52]{poisson2004}. Thus, Eq.\eqref{atlambda} becomes $\alpha dt/dr=$ constant, and since on null radial geodesics, $ds=0$ and $d\Omega=0$, we get from the metric that
\begin{equation}
\alpha(r)\beta(r)=\text{constant}.
\end{equation}
By normalization, this constant equals 1. Now, we can use the Raychaudhuri equation to determine $\alpha(r)$ and $\beta(r)$. The expansion parameter becomes
\begin{equation}
\theta=k^a_{~;a}=\frac{r \beta (r) \alpha'(r)+r \alpha (r) \beta '(r)+4 \alpha (r) \beta (r)}{2 r \alpha (r) \beta (r)},
\end{equation}
which simplifies to $\theta=2/r$ when $\alpha(r)\beta(r)=1$, and hence $d\theta/d\lambda+\theta^2/2=0$. Also $R_{ab}k^ak^b=0$, even though the Ricci tensor itself need not be zero.  Thus, The quantum Raychaudhuri equation simplifies considerably, and the terms that do not vanish are
\begin{equation}
\epsilon_1 \hbar^2 h^{ab} R_{;a;b} + \hbar^2 h^{ab} \left( \frac{\Box {\cal R}}{\cal R} \right)_{;a;b}~=0.
\end{equation}
The transverse metric $h^{ab}$ can be calculated from Eq.\eqref{auxmetric}, where an auxiliary null vector field that satisfies the conditions $k^aN_a=-1$ and $N^aN_a=0$ is
\begin{equation}
\label{Na}
N_t=\frac{-\alpha}{1+\sqrt{\alpha\beta}}, \quad N_r=\frac{-\sqrt{\alpha\beta}}{1+\sqrt{\alpha\beta}}, \quad N_{\theta}=N_{\phi}=0.
\end{equation}
For simplicity, we first consider the wave function $\psi={\cal R}e^{iS}$ with $\cal R =$ constant, and choose $S$ such that $k_a=\partial_aS$. Then, $\Box {\cal R}/{\cal R}=0$ and the quantum Raychaudhuri equation leads to the differential equation
\begin{equation}
r^3 \alpha'''(r)+4 r^2 \alpha''(r)-2 r \alpha '(r)-4 \alpha (r)+4=0,
\end{equation}
which has the solution
\begin{equation}
\alpha(r)=1+\frac{c_1}{r}+\frac{c_2}{r^2}+c_3 r^2.
\end{equation}
Asymptotic flatness requires $c_3=0$, and similarity with the standard metric when $\hbar\to0$ leads to $c_1=-2M$, and $c_2=\eta\hbar$ for some constant $\eta$. The constant $\eta$ must be dimensionless because $\hbar$ has dimensions of (length)$^2$ in geometric units. Thus, the modified Schwarzschild metric becomes
\begin{equation}
\label{modmetric}
ds^2=-\left(1-\frac{2M}{r}+\frac{\hbar\eta}{r^2}\right)dt^2+ \frac{1}{\left(1-\frac{2M}{r}+\frac{\hbar\eta}{r^2}\right)}dr^2+r^2d\Omega^2.
\end{equation}
From this metric, the Ricci scalar $R=0$ which means the intrinsic geometry is flat, but the Ricci tensor $R_{ab}$ is not zero, and the nonzero components are given by
\begin{align}
& R_{00}=\frac{\eta  \hbar  \left(\eta  \hbar -2 M r+r^2\right)}{r^6}, \qquad R_{11}=\frac{-\eta  \hbar }{r^2 \left(\eta  \hbar -2 M r+r^2\right)}  \\
& R_{22}=\frac{\eta  \hbar }{r^2}, \qquad R_{33}=\frac{\eta  \hbar  \sin^2\theta}{r^2}.
\end{align}
If we assume that the Einstein equations remain the same with the quantum corrections incorporated in the stress-energy tensor, then
\begin{equation}
R_{ab}-\frac{1}{2}g_{ab}R=8\pi T_{ab}
\end{equation}
leads to a stress-energy tensor with the non-zero components
\begin{align}
& T_{00}=\frac{\eta\hbar(r^2-2Mr+\eta\hbar)}{8\pi r^6}, \qquad  T_{11}=\frac{-\eta\hbar}{8\pi r^2(r^2-2Mr+\eta\hbar)},   \\\nonumber
& T_{22}=\frac{\eta\hbar}{8\pi r^2}, \qquad T_{33}=\frac{\eta\hbar\sin^2(\theta)}{8\pi r^2}.
\end{align}
The divergence of this tensor is zero $T^{ab}_{~~;b}=0$ which means the modified stress-energy tensor is also conserved. We note here the non-zero values of $R_{ab}$ and $T_{ab}$ which may be explained as a new source of energy due to the quantum correction that could correspond to dark energy and dark matter. The implications of these non-zero components need further investigations, and we hope to report on them soon. In the following section, we discuss how the new quantum corrections may ameliorate the physical singularity of the black hole.

\section{Black Hole Singularity}
For the modified metric \eqref{modmetric}, 
the Kretschmann scalar, which is given by contracting the Riemann tensor, is 
\begin{equation}
K=R_{abcd}R^{abcd}=\frac{8 \left(6 M^2 r^2-12 \eta\hbar Mr+7 \eta ^2 \hbar ^2 \right)}{r^8}.
\end{equation}
The Kretschmann scalar goes to infinity at $r=0$, which indicates an infinite curvature at $r=0$.  However, as argued in \cite{Das:2013oda}, the deviation equation does not equal zero at all times, and thus the quantal trajectories never go there and feel the singularity. To show this explicitly for Schwarzschild black hole, consider a reference frame of an observer falling from rest towards the black hole. A suitable coordinates for that frame is the Painlev\'{e}-Gullstrand coordinates introduced independently by Painlev\'{e} \cite{painleve} and Gullstrand \cite{gullstrand} in 1921. In these coordinates, the time coordinate is the proper time of a free-falling observer starting from infinity with zero velocity, and a slice of spacetime at fixed time corresponds to flat space. 
To derive\footnote{This derivation follows ref. \cite[P.417]{zee2013}. A more rigorous derivation and generalizations of these coordinates can be found in \cite{Martel:2000rn}.} the metric in these coordinates, we start by defining a new time coordinate
\begin{equation}
dT=dt+h(r)dr,
\end{equation}
with some function $h(r)$. Substituting in the metric we get
\begin{equation}
ds^2=-\alpha(r)dT^2+2h(r)\alpha(r)dTdr+\left(\frac{1}{\alpha(r)}-\alpha(r)h^2(r)	\right)dr^2+r^2d\Omega
\end{equation}
where $\alpha(r)=1-2M/r+\eta\hbar/r^2$. Choosing the coefficient of $dr^2$ to equal 1 fixes $h(r)=\frac{\sqrt{1-\alpha(r)}}{\alpha(r)}$, and leads to the metric
\begin{equation}
\label{rainmet}
ds^2=-\alpha(r)dT^2+2\sqrt{\frac{2M}{r}-\frac{\eta\hbar}{r^2}~}dTdr+dr^2+r^2d\Omega,
\end{equation}
which does not have a coordinate singularity at the horizon.

This metric can be written as
\begin{equation}
\label{rainmet2}
ds^2=\left[dr+\left(1+\sqrt{\frac{2M}{r}-\frac{\eta\hbar}{r^2}~}\right)dT\right] \left[dr-\left(1-\sqrt{\frac{2M}{r}-\frac{\eta\hbar}{r^2}~}\right)dT\right] +r^2d\Omega.
\end{equation}
For light on radial geodesic ($ds=0, d\Omega=0$), Eq.\eqref{rainmet2} has two solutions
\begin{equation}
\label{lightv}
\frac{dr}{dT}=-\sqrt{\frac{2M}{r}-\frac{\eta\hbar}{r^2}~} \pm 1.
\end{equation}
The positive sign represents light moving in the outward direction, while the negative sign represents light moving in the inward direction.

\begin{figure}[t]
	\centering
	\begin{minipage}[b]{0.48\linewidth}
		\includegraphics[width=\linewidth]{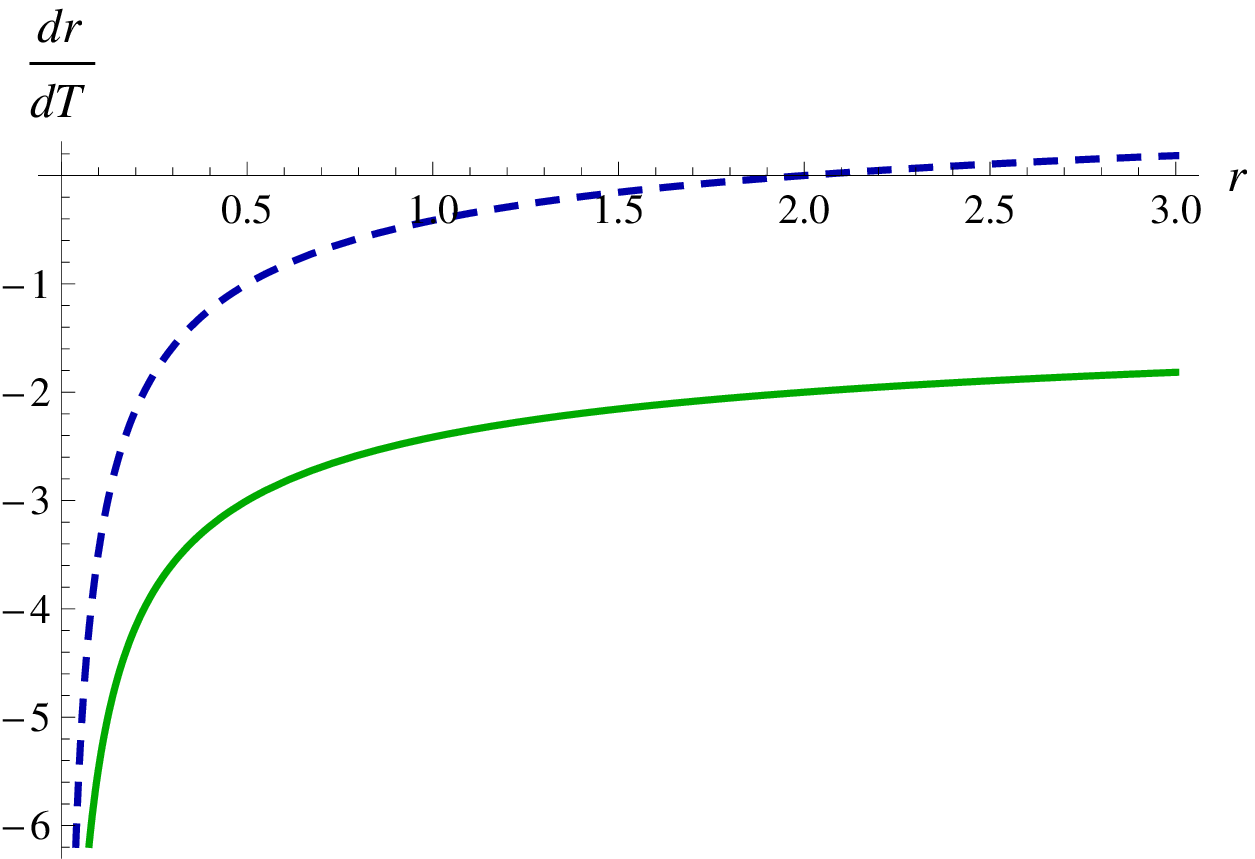}
		\caption{\label{fig:lightstd} Standard result for the velocity of light observed in a free-falling frame. Solid curve for light moving inward, and dashed curve for light moving outward. In this plot $M=1$.}
	\end{minipage}
	\quad
	\begin{minipage}[b]{0.48\linewidth}
		\includegraphics[width=\linewidth]{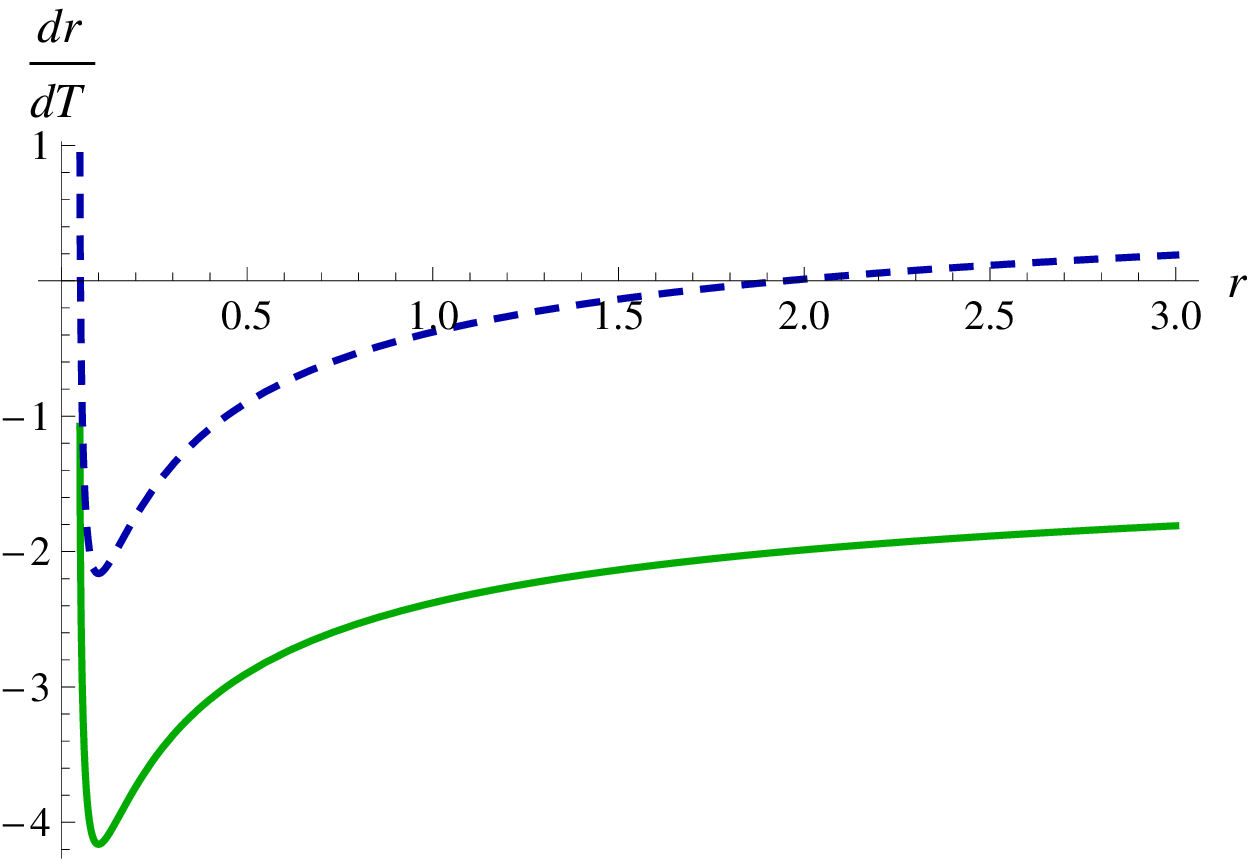}
		\caption{\label{fig:lightqre}
		The velocity of light observed in a free-falling frame according to Eq.\eqref{lightv}. Solid curve for light moving inward, and dashed curve for light moving outward. In this plot $\eta=1, \hbar=0.1$, and $M=1$}
	\end{minipage}
\end{figure}

Fig.\ref{fig:lightstd} is a plot of the standard result for the velocities of light  $\frac{dr}{dT}=-\sqrt{\frac{2M}{r}} \pm 1$. The velocities of both the inward and outward light rays go to negative infinity as $r\to0$, which means that anything that enters the event horizon reaches the singularity at the center.

Fig.\ref{fig:lightqre} is a plot of the modified velocity of light in Eq.\eqref{lightv}. The velocity reaches a minimum value at $r=\eta\hbar/M$ but then increases for smaller values of $r$. 
At $r=\eta\hbar/2M$, the velocity of an inward light ray reaches $-1$, and the velocity of an outward light ray reaches $1$.
Below this value, $dr/dT$ is imaginary, which means that nothing can reach the center, and the observed curvature scalars remain finite.

This behavior of free-falling particles is similar to that for the Reissner-Nordstr\"{o}m black hole since the Reissner-Nordstr\"{o}m metric is similar to the metric \eqref{modmetric} with $Q^2 \to \eta\hbar$. Between the outer horizon $r_+$ and the inner Cauchy horizon $r_-$, the coordinate $r$ becomes timelike and a falling particle must continue inwards. However, below $r=r_-$, the coordinate $r$ is spacelike and a falling particle can move in the direction of increasing $r$ until $r=r_-$ where $r$ becomes timelike again but in the reverse direction, i.e. the particle must move towards increasing $r$. So the particle gets into a wormhole at the Cauchy horizon and gets out of a white hole in a different universe. (See the Penrose diagram in \cite[P. 921]{misner1973} or \cite[P. 258]{carroll2004})However, in physically realistic situations, a signal approaching $r_-$ gets infinitely blue-shifted, which  disturbs the inner geometry of the black hole making the Cauchy horizon unstable \cite{chandrasekhar1982, poisson1989inner}. In \cite{poisson1989inner, poisson1990internal}, it was shown that, when perturbed, the inner horizon of a charged black hole becomes a singularity with infinite spacetime curvature (the mass-inflation singularity). Thus, there is good reason to believe that the Cauchy horizon is a true physical singularity. For the modified Schwarzschild metric we found a Cauchy horizon of Planckian size, so perhaps it is reasonable to speculate that this Cauchy horizon is actually the radius of the black hole singularity, i.e. a non-point like singularity.

For the standard Schwarzschild metric, the singularity is spacelike, but for the quantum-corrected metric, the singularity is timelike. This could have implications for the information loss problem. In Reissner-Nordstr\"{o}m and Kerr black holes, the singularity is timelike and the Penrose diagram is an infinite set of copies, so information could come out of the singularity and reach the future null infinity of another universe. (Perhaps, due to quantum corrections, signals can come out to our universe and hence there is not information loss.) For a spacelike singularity, this seems difficult/impossible, but there seems to be some hope for timelike singularities, so it is interesting that due to quantum corrections, the spacelike singularity of Schwarzschild black hole becomes timelike.
In the next section, we investigate the implications of the quantum corrections on black hole thermodynamics.

\section{Modified Black Hole Thermodynamics}

\begin{figure}[t]
	\centering
	\includegraphics[width=0.5\linewidth]{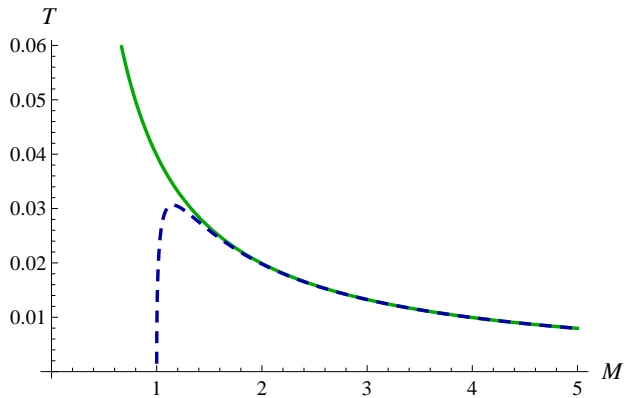}
	\caption{\label{fig:temp} Standard and modified temperature plotted against mass when $\eta=1$ and $\hbar=1$.}
\end{figure}

The modified black hole temperature from the metric \eqref{modmetric} can be calculated from the surface gravity by
\begin{equation}
T_H=\frac{\kappa}{2\pi}=\lim_{r\to r_+} \frac{1}{4\pi}\frac{dg_{tt}}{dr},
\end{equation}
where $r_+$ is the outer horizon radius, which is found from the metric by setting $1-\frac{2M}{r}+\frac{\hbar\eta}{r^2}=0$ to get
\begin{equation}
r_{\pm}=M\pm \sqrt{M^2-\eta\hbar}.
\end{equation}
Thus, the modified temperature is given by
\begin{equation}
T=\frac{\sqrt{M^2-\eta\hbar }}{2 \pi  \left(\sqrt{M^2-\eta  \hbar }+M\right)^2},
\end{equation}
which goes to zero at $M=\sqrt{\eta\hbar}$  signaling the existence of a remnant. Figure \ref{fig:temp} shows a plot of the modified and standard temperature. In other words, the Hawking BH temperature is vanishing (no Hawking radiation). Therefore, our quantum corrections lead to forming  an extremal BH \cite{Nicolini:2005vd} which yields  a ``frozen" horizon. In the Bonanno-Reuter  model, $T_H = 0$ means the BH evaporation stops and the mass $M$ is critical (a soliton-like remnant is formed \cite{Bonanno:2000ep}). 

\begin{figure}[t]
	\centering
	\begin{minipage}[b]{0.48\linewidth}
		\includegraphics[width=\linewidth]{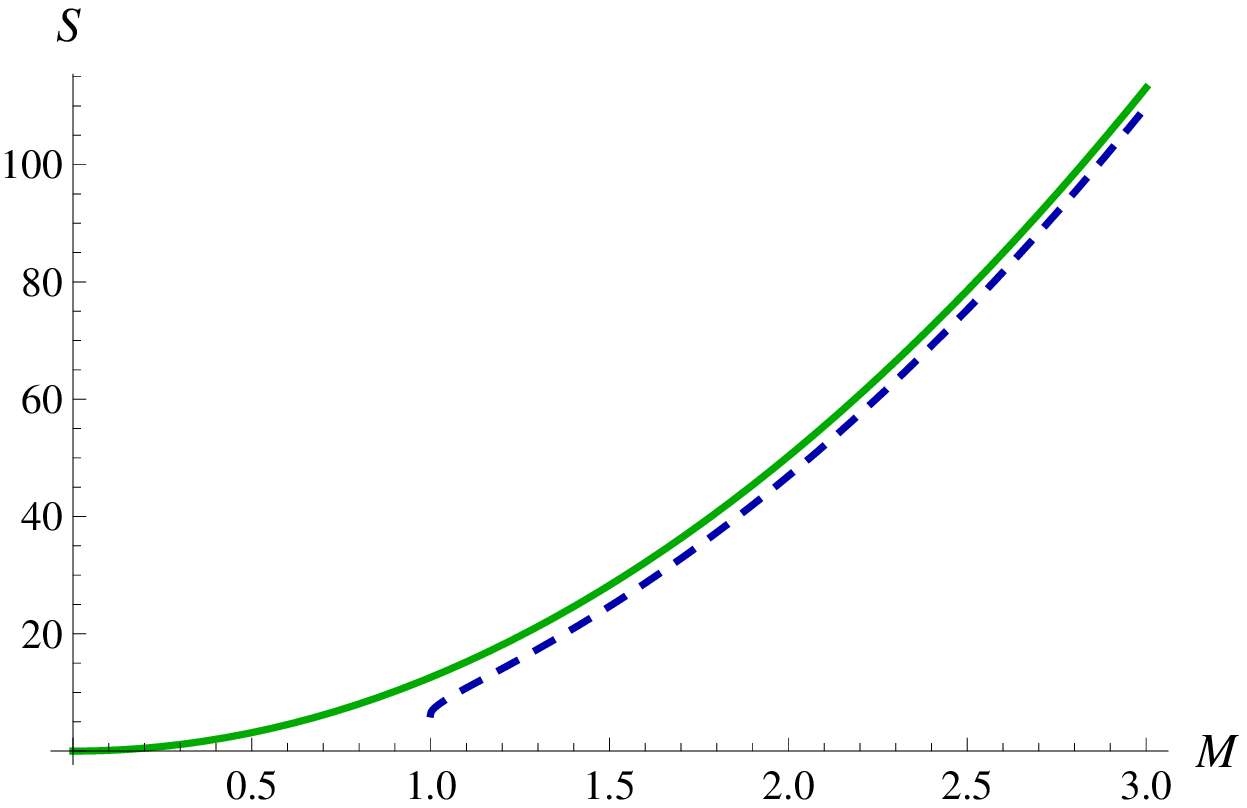}
		\caption{\label{fig:entropy} Standard and modified entropy plotted against mass assuming $\eta=1$ and $\hbar=1$.}
	\end{minipage}
	\quad
	\begin{minipage}[b]{0.48\linewidth}
		\includegraphics[width=\linewidth]{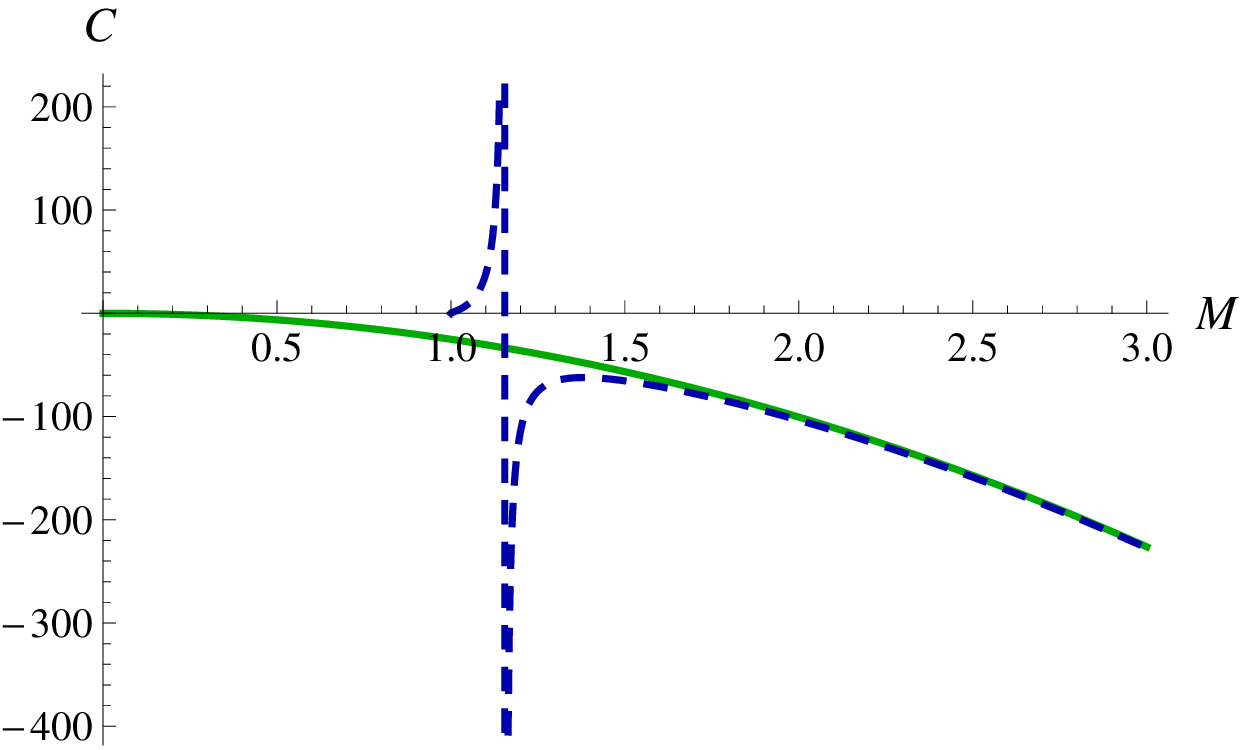}
		\caption{\label{fig:capacity}
			Standard and modified heat capacity plotted against mass assuming $\eta=1$ and $\hbar=1$.}
	\end{minipage}
\end{figure}

The entropy can be calculated from the first law of black hole mechanics $dM=TdS$
\begin{equation}
S=\int \frac{dM}{T}=2 \pi  \left(M \sqrt{M^2-\eta  \hbar }+M^2\right),
\end{equation}
which is undefined when $M$ goes below the value $M=\sqrt{\eta\hbar}$ as can be seen from figure \ref{fig:entropy}.  The heat capacity can be calculated from the relation
\begin{equation}
C=T\frac{\partial S}{\partial T}=-\frac{2 \pi  \sqrt{M^2-\eta  \hbar } \left(\sqrt{M^2-\eta  \hbar }+M\right)^2}{2 \sqrt{M^2-\eta  \hbar } - M},
\end{equation} 
which again goes to zero at $M=\sqrt{\eta\hbar}$, which means that the black hole does not exchange heat with the surrounding space. Figure \ref{fig:capacity} is a plot of the standard and modified heat capacity and it diverges at the value of maximum temperature.

These results mean that the quantum corrections prevents BHs from evaporating completely, just like  quantum mechanics  prevents the hydrogen atom from collapsing \cite{Adler:2001vs}. Since a black hole remnant possess a frozen horizon, we think that our result may ameliorate the information loss paradox.

\section{Time-Dependent Modified Metric}

\begin{figure}[t]
	\centering
	\includegraphics[width=0.5\linewidth]{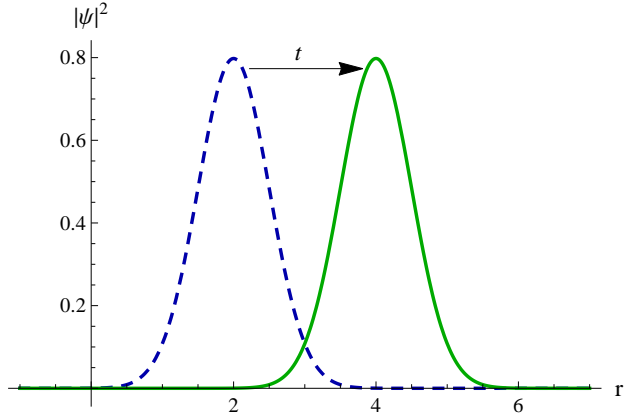}
	\caption{\label{fig:psi}  $|\psi|^2$ as $t$ increases from $t=2$ (left) to $t=5$ (right) assuming $\sigma=1$}
\end{figure}

In Sec.\ref{sec:metric}, we derived the metric for a wave function $\psi={\cal R}e^{iS}$ with constant $\cal R$. In this section, we consider a less trivial example for the wave function. Consider a wave packet moving at the speed of light on an outgoing radial null geodesic. Such a wave function can be represented by
\begin{equation}
\label{psi}
\psi=A e^{-(r-t)^2/\sigma}e^{iS},
\end{equation}
where $\sigma$ is the spread of the wave packet, $A=(2/\sigma^2\pi)^{1/4}$ is a normalization factor, and $S$ is chosen such that $k_a=\partial_a S$. Figure \ref{fig:psi} shows $|\psi|^2$ at two different times, and we see it moves outwards as $t$ increases. Since the wave function is time-dependent, we expect the metric to  also be time-dependent. So we start with the standard Schwarzschild metric after adding a time-dependent function $f(r,t)$ and then use the quantum Raychaudhuri equation to find that function. We begin with the metric ansatz
\begin{equation}
ds^2=-\alpha(t,r)dt^2+\frac{1}{\alpha(t,r)}dr^2+r^2 d\Omega^2,
\label{metrict}
\end{equation}
with
\begin{equation}
\alpha(t,r)=1-\frac{2M}{r}+\hbar f(t,r).
\end{equation}
For the wave function \eqref{psi} we get
\begin{equation}
\frac{\square {\cal R}}{\cal R}=\frac{4(r-t)}{r\sigma^2}.
\end{equation}

Using $k_a$ and $N_a$ from Eqs.\eqref{ka} and \eqref{Na} but with  time-dependent $\alpha(t,r)$  we get that
\begin{equation}
h^{ab}\left(\frac{\square {\cal R}}{\cal R}\right)_{;a;b}= \frac{8 t (r \hbar  f(t,r)-2 M+r)}{r^4 \sigma ^2}.
\end{equation}
Now, the quantum Raychaudhuri equation to first order in $\hbar$ leads to the differential equation
\begin{equation}
(2 M r-r^2) \frac{\partial^2 f(t,r)}{\partial t\partial r}+12 M \frac{\partial f(t,r)}{\partial t}-4 r \frac{\partial f(t,r)}{\partial t}=0,
\end{equation}
which has the solution
\begin{equation}
f(r,t)=\eta  \left(\frac{2 M^2 t^2}{r^6}-\frac{2 M^2}{r^6}-\frac{2 M t^2}{r^5}+\frac{2 M}{r^5}+\frac{t^2}{2 r^4}-\frac{1}{2 r^4}\right),
\end{equation}
for some constant $\eta$, and the metric is modified by
\begin{equation}
\alpha(t,r)=1-\frac{2M}{r}+\eta\hbar\left(\frac{t^2}{2 r^4}-\frac{1}{2 r^4}-\frac{2 M t^2}{r^5}+\frac{2 M}{r^5}+\frac{2 M^2 t^2}{r^6}-\frac{2 M^2}{r^6}\right).
\end{equation}
If we apply this metric to a particle in orbit around a black hole, we see that the quantum correction term grows as time passes and goes to infinity as $t\to\infty$, which of course is unphysical. This is because the metric was derived for a wave function describing a particle on an outgoing radial geodesic, and on that geodesic $O(t)\sim O(r)$. Hence, the metric goes to zero as $r\to\infty$. So this metric describes only a particle on a radial geodesic, but cannot be generalized to other geodesics. Thus, since the quantum Raychaudhuri equation depends on the wave function of the particle under study, the resulting modified metric will also depend on that wave function and will be different for different particles and geodesics.

\section{Constraints on $\eta$}

In Sec. \ref{sec:metric}, we determined the metric \eqref{modmetric} up to a constant $\eta$. The value of that constant can be constrained from astrophysical observations and laboratory experiments such as deflection of light, and gravitational redshift.

\subsection{Deflection of Light}
When light approaches a massive body, such as the sun, it gets deflected from a straight line by an angle given by \cite[P.189]{weinberg}
\begin{equation}
\label{deflection}
\Delta\phi=2\int_{r_0}^{\infty}\frac{1}{r\sqrt{\alpha(r)}}\left(\frac{r^2}{r_0^2}\frac{\alpha(r_0)}{\alpha(r)}-1 \right)^{-1/2}dr-\pi,
\end{equation}
where $r_0$ is the distance of closest approach to the sun. In general relativity the deflection angle to first order in $M/r_0$ is given by
\begin{equation}
\label{grdefl}
\Delta\phi_{GR}\simeq \frac{4M}{r_0}.
\end{equation}
To calculate that deflection angle for the modified metric, we begin by making the transformation $u\equiv r_0/r$ in Eq. \eqref{deflection}
\begin{equation}
\label{deflectionu}
\Delta\phi=2\int_{0}^{1}\frac{1}{\sqrt{\alpha\left(\frac{r_0}{u}\right)}}\left(\frac{\alpha(r_0)}{\alpha\left(\frac{r_0}{u}\right)}-u^2 \right)^{-1/2}du-\pi.
\end{equation}
To simplify the integral, we expand the integrand to first order in $\hbar$ and $M/r_0$ to get
\begin{equation}
\frac{1}{\sqrt{\alpha\left(\frac{r_0}{u}\right)}}\left(\frac{\alpha(r_0)}{\alpha\left(\frac{r_0}{u}\right)}-u^2 \right)^{-1/2}  \approx 
\frac{1}{\sqrt{1-u^2}}+\frac{M \left(u^2+u+1\right)}{r_0 (u+1) \sqrt{1-u^2}}-\frac{\eta  \left(u^2+1\right) \hbar }{2 r_0^2 \sqrt{1-u^2}}.
\end{equation}
Thus, Eq. \eqref{deflectionu} evaluates to
\begin{equation}
\Delta\phi \approx \frac{4M}{r_0}-\frac{3 \pi  \eta  \hbar }{4 r_0^2}.
\end{equation}
The best accuracy of measuring the deflection of light by the sun is from measuring the deflection of radio waves from distant quasars using the Very Long Baseline Array (VLBA) \cite{Fomalont:2009zg}, which achieved an accuracy of $3\times 10^{-4}$. Thus,
\begin{equation}
\frac{\delta\Delta\phi}{\Delta\phi_{GR}} \approx \frac{3\pi\eta\hbar/4r_0^2}{4M/r_0} < 3\times 10^{-4}.
\end{equation}
Assuming that light grazes the surface of the sun $r_0\simeq R_\odot=6.96\times 10^8\text{m}$ and $M=M_\odot=1.99\times 10^{30}\text{kg}=1.477\times10^{3}\text{m}$ with $\hbar=2.61\times10^{-70} \text{m}^2$ in geometric units, we get an upper bound on $\eta$ of
\begin{equation}
\eta< 2.0\times 10^{78}.
\end{equation}

\subsection{Gravitational Redshift}
When light moves between two radii $r_1$ to $r_2$, its frequency changes. To find the frequency change, we apply the metric at the two radii by putting $dr=0$ and $d\phi=0$, and solve for $dt$. Since the time measured by a remote observer is the same for the two radii, we get
\begin{equation}
	dt=\frac{d\tau_1}{\sqrt{\alpha(r_1)}}=\frac{d\tau_2}{\sqrt{\alpha(r_2)}},
\end{equation}
where $d\tau^2=-ds^2$, and $\alpha=1-2M/r+\eta\hbar/r^2$. The relative frequency is
\begin{equation}
	\frac{\omega_2}{\omega _1}=\frac{d\tau_1}{d\tau _2}=\sqrt{\frac{\alpha(r_1)}{\alpha(r_2)}}.
\end{equation}
expanding to first order in $\hbar$
\begin{equation}
	\frac{\omega_2}{\omega _1}=\sqrt{\frac{1-2M/r_1}{1-2M/r_2}} \left(1+\frac{r_2^2-r_1^2-2Mr_2+2Mr_1}{2r_1 r_2(2M-r_1)(2M-r_2)}\eta\hbar\right).
\end{equation}
Subtracting one from the previous result we get
\begin{align}
	\label{redshift}
	\frac{\omega_2-\omega_1}{\omega_1}=(S-1)
	\left(1+\left(\frac{S}{S-1}\right)\frac{r_2^2-r_1^2-2Mr_2+2Mr_1}{2r_1 r_2(2M-r_1)(2M-r_2)}\eta\hbar\right),
\end{align}
where
\begin{equation}
	S\equiv\sqrt{\frac{1-2M/r_1}{1-2M/r_2}}.
\end{equation}
The most accurate measurement of gravitational redshift is from Gravity Probe A \cite{Vessot:1980zz} in 1976. The satellite was at an altitude of $10^7$m, and achieved an accuracy of $7.0\times 10^{-5}$, which means that the new term in Eq. \eqref{redshift}
\begin{equation}
	\frac{S}{S-1}\frac{r_2^2-r_1^2-2Mr_2+2Mr_1}{2r_1 r_2(2M-r_1)(2M-r_2)}\eta\hbar<7.0\times 10^{-5}.
\end{equation}
Using the mass and radius of the earth, $M_\oplus=5.97\times10^{24}\text{kg}=4.44\times10^{-3}\text{m}$, $ r_1=R_\oplus=6.38\times10^{6}\text{m}$, and $r_2=r_1+10^7 \text{m}$, we get the bound
\begin{equation}
	\eta < 1.1\times 10^{70}.
\end{equation}
This bound is large, but not inconsistent with observations. Also, since the constant $\eta$ could be much larger than one, future experiments could produce better constraints.

\section{Conclusions}

In this work, we investigated the implications of quantum corrections, that have been obtained by replacing quantal geodesics with classical geodesics, on  black hole physics. We obtained a new modified Schwarzschild metric, and found that the thermodynamic properties of this quantum black hole are greatly changed when the black hole size approaches the Planck scale. We found that
our quantum corrections lead to forming  an extremal BH which yields  a ``frozen" horizon and hence the BH evaporation stops, forming a black hole remnant.  Besides, we found that the physical singularity of the black hole turned to be a timelike singularity, and that the quantal trajectories never go there and feel the singularity. Timelike singularity  has the interesting feature that a timelike curve can be constructed such that an observer can have the singularity next to them without necessarily falling into it. This may ameliorate the information loss problem.
These results need further studies with different types of black holes. We hope to report on these in the future.

\begin{acknowledgments}
	The research of AFA is supported by Benha University (www.bu.edu.eg). The authors would like to thank Saurya Das for constructive comments and suggestions that significantly helped in writing this paper.
\end{acknowledgments}

\bibliography{refs}

\end{document}